# Analysis of a spinning polygon wavelength swept laser

Bart Johnson, Walid Atia, Mark Kuznetsov, Brian D. Goldberg,
Peter Whitney, and Dale C. Flanders

*Axsun Technologies, 1 Fortune Drive, Billerica, MA 01821, USA*

**Abstract:** It has been known for quite some time that spinning polygon, and similar, swept lasers used in OCT favor the short to long wavelength sweep direction because of four wave mixing in the gain medium. Here we have reformulated the problem in the time domain and show experimentally and through numerical simulation that these lasers are pulsed. The emitted pulses modulate the gain medium refractive index to red shift the light. Instead of new wavelengths being built up slowly from spontaneous emission, each pulse hops to a longer wavelength by nonlinear means, tracking the tunable filter. This allows high speed, low noise tuning in the blue to red direction. Based on this model, we make the first coherence length calculations for a swept source.

## 1. Introduction

One of the biggest transformations in OCT was the recognition that frequency-domain methods were faster, for the same signal to noise ratio, than time-domain methods [1-6]. This immediately brought about an interest in rapidly swept tunable laser sources [7]. Some of these sources [8,9] showed asymmetric tuning behavior where more power was emitted when swept in the short-to-long wavelength direction. This was attributed to four-wave mixing in the semiconductor optical amplifier (SOA) gain medium [8,9]. Later, a computer simulation of these lasers [10] confirmed the mechanism behind the asymmetric tuning. The simulations also showed a variety of laser behaviors, including optical pulsation, for a relatively long fiber optic ring laser cavity. They also showed the importance of the linewidth enhancement factor [11] in producing these behaviors. We [12] and other groups have simulated rapidly swept short cavity lasers [13,14], showing tuning direction asymmetry, pulsation, and in some cases, mode locking. The short-cavity swept laser described in [12,15] is passively mode locked, and can also be stabilized by active mode locking through synchronous modulation of the injection current [16].

   This paper starts with a brief physical description of how pulsation aids rapid wavelength tuning in the short to long wavelength direction. Numeric simulation results, based on the mode locking theories of Haus [17,18], are presented. His theories are actually general laser models that are used to find mode locked solutions. Without approximation, they are capable of predicting laser behavior that is not necessarily mode locked. We use Haus as a starting point and add the effect of the imaginary part of the gain and a relatively narrow dynamically tuned filter. We then compare the simulations to experimental behavior of a commercial Santec HSL-2000 spinning-polygon-type laser that sweeps 100 nm at 20 kHz in the 1310 nm wavelength range. Similar designs are described in [19-21].

## 2. Laser nonlinear tuning mechanism

Rapidly swept lasers under consideration here tune too quickly for lasing to build up anew from spontaneous emission at each new wavelength [22]. In these cases, a nonlinear optical mechanism is required to shift the wavelength of light circulating within the laser cavity to match the wavelength of the filter on successive round trips. With the exception of specially designed lasers [20,21] and tunable VCSELs [23,24], Doppler shift from the moving optical tunable filter mechanism only does part of the job. Most of the shift comes from phase modulation induced by depletion of the gain as the optical pulses travel through the semiconductor optical amplifier. Gain depletion is accompanied by a rise in refractive index of the gain medium. The coupling between the index and the power gain can be described using the linewidth enhancement factor [11], $\alpha$, as:

$$\Delta n = -\alpha \frac{\lambda}{4\pi} \Delta g \qquad (1)$$

The pulsation process, for a short cavity laser that emits one pulse per round trip, is illustrated in Fig. 1. The SOA becomes optically longer as the pulse travels through, red shifting the light field. The laser does not tune continuously, but rather hops discretely to the next wavelength on each new pulse. The frequency hop for a SOA of length $L$ is:

$$\Delta \nu = -\frac{L}{\lambda} \frac{dn}{dt} \qquad (2)$$

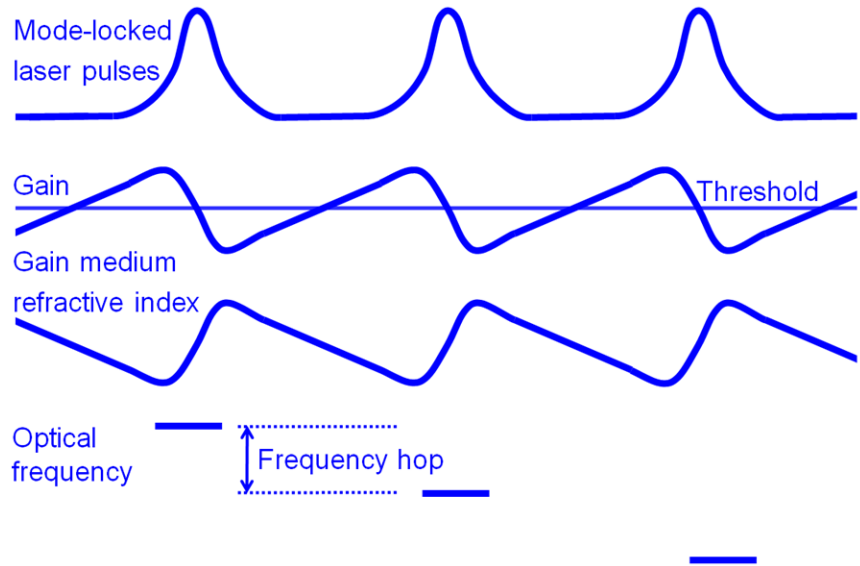

Fig. 1. Pulsation and frequency hopping dynamics of a short cavity rapidly swept laser.

The pulse energy and width determine the magnitude of the frequency hop. The laser operates in this manner because the lowest threshold is obtained when the pulse frequency hops to follow the filter tuning. A feedback mechanism built into the laser dynamics naturally ensures the pulse frequency hops to follow the filter.

**3. Experimental details**

The laser configuration is shown in Fig. 2, and data from two sweeps of the Santec HSL-2000 are presented in Fig. 3. Variations on this design are reported in [19-21]. The tuning rate varies across the sweep (Fig 3A), but is in the neighborhood of -0.5 GHz/ns. The pulsation is captured in Fig. 3C at a 10 GS/s sample rate. The detection bandwidth is limited by an oscilloscope nominally rated at 2.5 GHz bandwidth, but still has usable response up to 5 GHz. The bandwidth limitation only allows the fundamental pulsation harmonic to be captured, so the pulses appear sinusoidal. When starting a sweep, the laser locks onto a pulsation rate and maintains that through the sweep. The two adjacent sweeps shown here are an extreme pair, the first with 33 pulses per round trip and the second at 44. Most traces show pulse numbers between these values. A spectrogram (Fig. 3D) of the photodiode power trace (Fig. 3C) shows high RF energy at the pulsation rates of 3.2 GHz (left sweep) and 4.2 GHz (right sweep). The spectrogram shows beating between all lasing cavity modes. The first harmonic is at c/L, where L is optical length of the ring cavity. Note that the spectrogram traces are slightly tilted. This shows that the cavity lengthens slightly as the sweep progresses. This is due to the construction of the spinning polygon tuning mechanism and it is partially responsible for the red shifting of the laser, contributing about -1.1 GHz of shift on each round trip (-0.1 GHz/ns average rate).

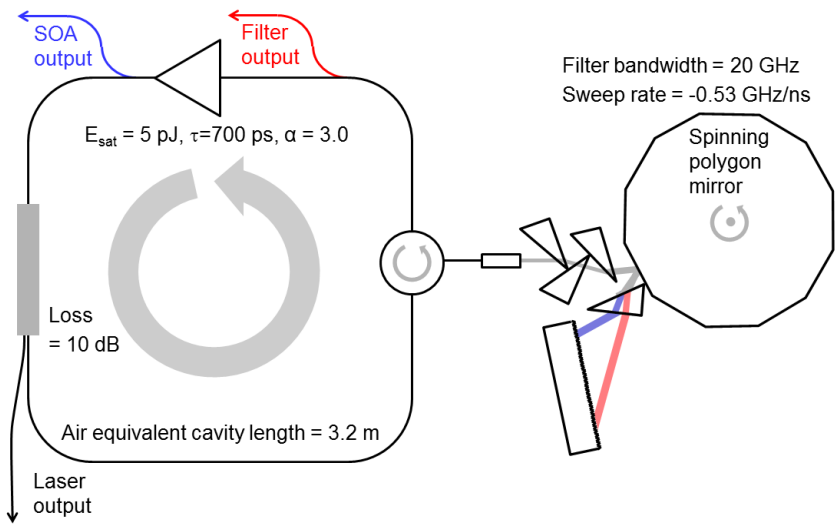

Fig. 2. Diagram of Santec HSL-2000 fiber ring laser with spinning polygon tuning mechanism.

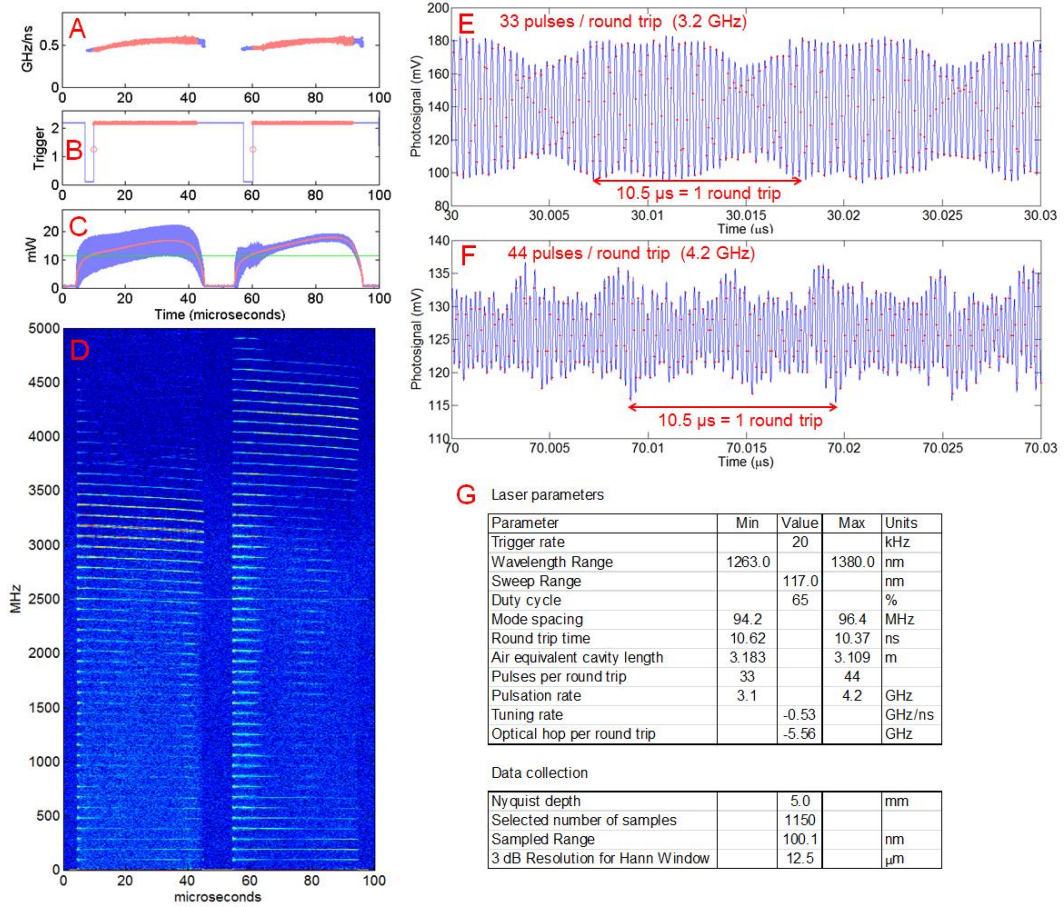

Fig. 3. Pulsation and frequency hopping dynamics of a Santec HSL-2000 swept laser.

## 4. Mathematical model of laser

Our laser model is a descendent of the mode-locking theories of Haus [17]. A detailed introduction to Haus' methods is provided in reference [18], where a number of mode locking problems are attacked in a similar way. We follow his basic method, but add the effects of the imaginary part of the gain and add a tunable intracavity filter and a fixed frequency shift per cavity pass. Instead of developing a master equation and solving it analytically, we move directly to numerical solution. Haus' theories are actually general laser models that are refined to obtain mode-locked solutions. They can be used to model more general laser behavior, as we have done here.

The analytical laser model is presented in Fig. 4. The diagram also defines some of the variables used in the analysis.

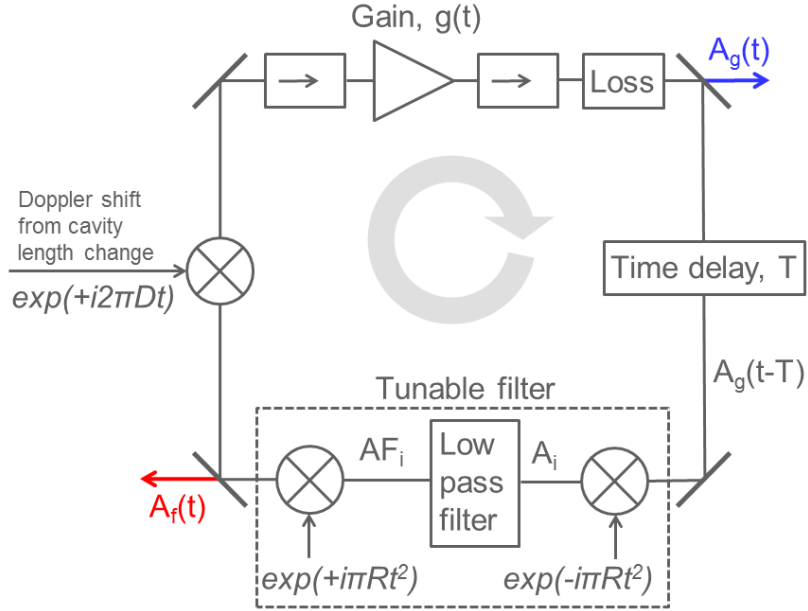

Fig. 4. Mathematical model of the swept laser.

The analysis starts with a rate equation for the dynamics of the semiconductor optical amplifier (SOA). $A_f(t)$ is the optical field delivered to the SOA from the swept optical filter; $g - g_0$ is the field gain integrated over the length of the SOA (the power gain is $2(g - g_0)$); $p$ is the pump level; $\tau$ the carrier lifetime; $\ell$ the fixed cavity loss; $E$ the SOA saturation energy. The pump, $p$, is normalized so that $p = 1$ at threshold for a zero tuning rate ($R = 0$).

$$\frac{dg}{dt} = p\frac{\ell}{\tau} - \frac{g}{\tau} - \left(\exp\left[2(g - g_0)\right] - 1\right)\frac{|A_f(t)|^2}{E} \tag{3}$$

The tunable filter is swept mathematically by mixing the light to baseband, low pass filtering, and then mixing back up to lightwave frequencies as shown in Fig. 4. It is done this way to separate the linear and nonlinear parts of the problem of sweeping the filter. This method naturally accounts for the Doppler shift produced by the moving mechanism of any swept filter. The method makes an exact accounting of the nonlinear frequency shift, but the magnitude of the effect is roughly $RT_G$ where $R$ is the tuning rate and $T_G$ is the filter's group delay. The wave is mixed down, but delayed in the low pass filter before being mixed back up again. That will produce a frequency shift. The optical tuning rate $R$ is defined in Fig. 4. The exact optical carrier frequency does not matter in the simulation. Frequencies lower than "lightwave" are actually used to prevent aliasing for numerical integration steps on the order of 1 ps.

The impulse response of the tunable filter, translated to DC, is modeled as a step exponential as described in Eq. (4) below, where $u(t)$ is the unit step function and $B$ is the FWHM bandwidth of this Lorentzian filter.

$$h(t) = \frac{1}{\pi B} \exp(-\pi B t) u(t) \tag{4}$$

The computer code implements this as an infinite impulse response (IIR) filter [26] for efficiency. The simulation variables $AF_{i-1}$ and $A_i$ are defined in Fig. 4.

$$AF_i = C_1 A_i + C_2 AF_{i-1} \tag{5}$$

For an integration time step $\Delta t$, the IIR filter coefficients are:

$$C_1 = 1 - \exp(-\pi B \Delta t) \tag{6}$$

$$C_2 = \exp(-\pi B \Delta t) \tag{7}$$

Finally, the field out of the SOA is linked to the field out of the swept filter, where $\ell$ is the fixed cavity loss and $\alpha$ is the linewidth enhancement factor.

$$A_g(t) = A_f(t) \exp\left[(1 + i\alpha)(g(t) - g_0) - \ell\right] \tag{8}$$

This model can predict a number of behaviors. For very short cavities, regularly mode-locked pulses are possible. The model does not assume regular pulsation, however. It is a general laser model and other (chaotic) solutions are possible, such as is the case when positive tuning rates (red-to-blue tuning) are applied. We will concentrate on the solutions that apply to the operational state of the Santec HSL-2000, which are more complex and involve many pulses per cavity round trip.

## 5. Pulsed solutions to the laser model

A numerical solution to the laser model showing pulsed behavior is shown in Fig. 5 for the parameters listed in Table 1, which correspond to a 1310 nm laser swept 100 nm at 20 kHz with a 64% duty cycle. These parameters are based on measurements, general estimates from semiconductor laser physics, and information on similar laser designs from the literature [19-21]. The first cavity transit is seeded with noise and then the model is run for 100's of nanoseconds at which point it achieves a dynamic steady state. No spontaneous emission noise is added during the simulation. Any "randomness" in the solution is the result of chaos, not spontaneous emission noise. In this example, the laser pulses 42 times per round trip, strongly modulating the gain. Fig. 5 shows the solution after 500 ns of evolution. Inside the mechanics of the calculation, the SOA output is time-aligned with the filter output, and the SOA responds to input from the filter, as seen in Figs. 2 and 4. In Fig. 5, the SOA solution is advanced by one round trip in the plot so that the red curve is the filtered output of the blue SOA curve instead of the other way around.

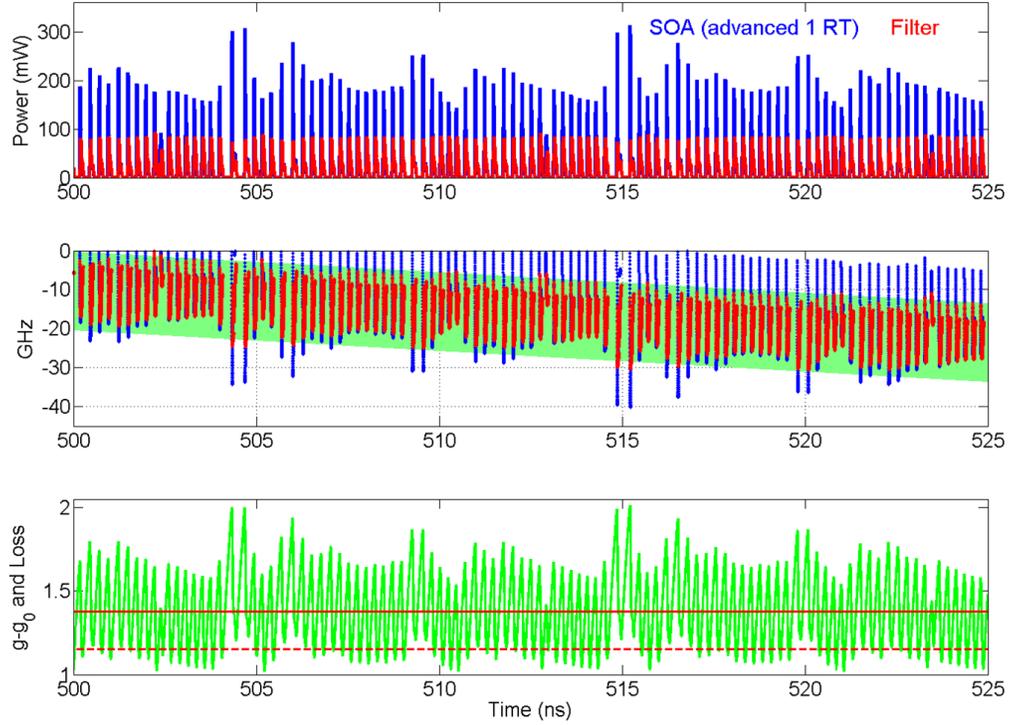

Fig. 5. Pulsed solutions to the swept laser model after 500 ns of evolution. Blue curves are derived from $Ag$ (SOA output) and red from $Af$ (filter output). The top plot shows the pulse powers, middle the instantaneous frequencies (minus an arbitrary fixed lightwave frequency), the lowest plot is the gain, $g - g_0$, in green, and the dynamic threshold (solid red = fixed losses plus dynamic filter loss) and the unswept ($R = 0$) threshold (dashed red = fixed losses only).

Table 1. Laser simulation parameters

| Parameter | Variable | Value |
|---|---|---|
| Sweep rate | $R$ | -0.53 GHz/ns |
| Doppler shift per cavity transit | $D$ | -1.1 GHz |
| Cavity delay | $T$ | 10.5 ns |
| Filter bandwidth | $B$ | 20 GHz |
| Normalized pump | $p$ | 4.0 |
| Linewidth enhancement factor | $\alpha$ | 3.0 |
| Carrier lifetime | $\tau$ | 700 ps |
| Loss | $\ell$ | 1.15 |
| Saturation energy | $E$ | 5 pJ |

A more detailed look at the pulses is taken in Fig. 6. Again, the SOA solution is plotted advanced by one round trip. The computer model fills in the details of the process outlined in Fig. 1. Each pulse "hops" to a new optical frequency, but there is also considerable chirp to the pulses. Freshly filtered pulses are longer and have smaller chirp than those that have just passed through the SOA. The filter also has a peak power leveling effect, throwing away the highly chirped portions of the SOA output pulses. This is because large pulses, such as the two between 504 and 505 ns, strongly modulate the SOA and chip themselves outside the filter bandwidth (shown by the green band in the middle plot). The portion of the pulse outside the filter bandwidth is strongly attenuated. The blue pulses lead the red ones slightly for two reasons: (1) The SOA preferentially builds up the front of the pulse before the gain is depleted, and (2) the group delay of the filter, approximately $1/(\pi B)$, delays the filter output by about 20 ps.

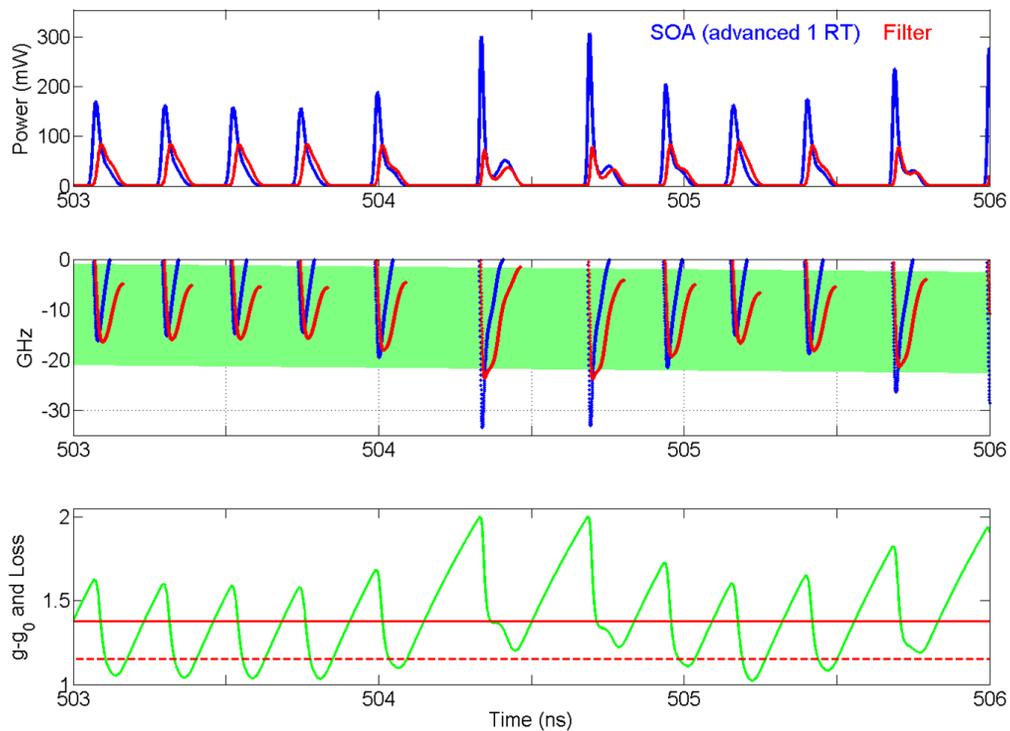

Fig. 6. Pulsed solutions to the swept laser model after approximately 500 ns of evolution with finer time resolution.

The solution evolves to one of more uniform pulsation in spacing and amplitude, but still with some residual modulation on every round trip of the cavity as shown in Fig. 7. Similar behavior is seen in the spectrogram (Fig. 3D) for the experimental laser. It shows that near the turn-on point for the sweep, there are many harmonics excited from very "random" behavior. After a brief period, ~2 µs in the left sweep and ~10 µs for the right, the laser settles into a steadier, smoother, more narrow band modulation at ~3.2 GHz (left trace) and ~4.2 GHz (right trace). This more uniform behavior is also predicted by the model. After 3 µs, the solution evolves from that of Figs. 5 and 6 to that in Fig. 7. The pulses are more uniform in spacing and amplitude and chirp does not stray far out of the filter bandwidth. The SOA curve is not advanced by one round trip in Fig. 7 so that the blue curve shows how much negative frequency shift is added to the light from the filter. It shifts the light by approximately the "hop" needed to match the filter position after one round trip. Any excess hop is trimmed off by the filter.

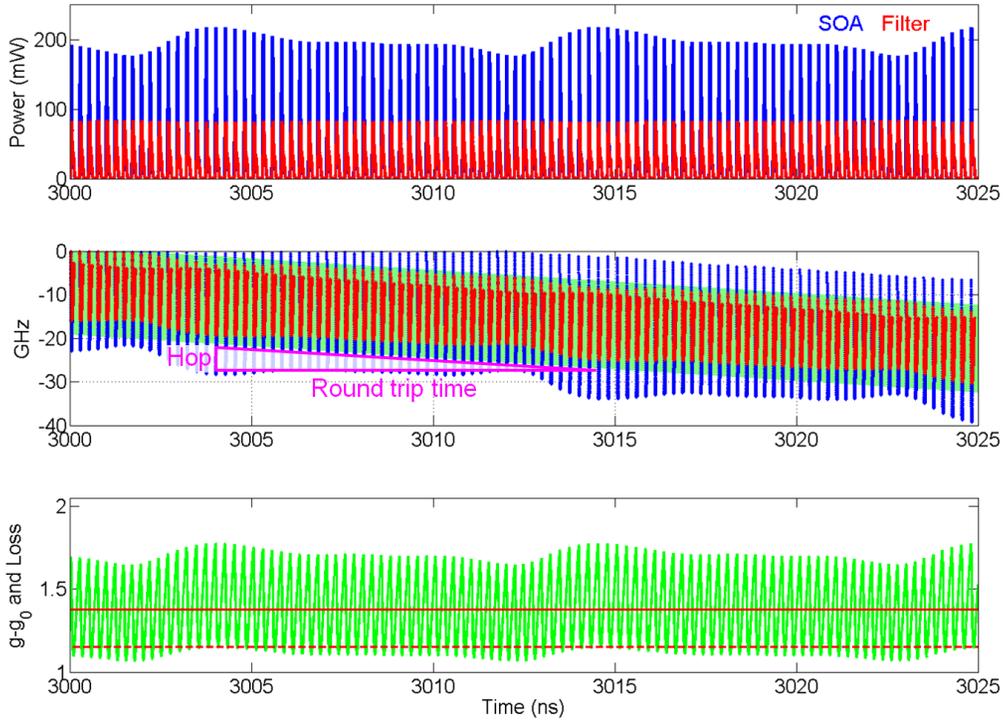

Fig. 7. Pulsed solutions to the swept laser model after 3000 ns of evolution.

## 6. Coherence properties

Output from the model is the light field's magnitude and phase, a complete description of the laser emission. Any experiment that can be performed in the laboratory can be simulated with these results. For example, Fig. 8 shows the results of a coherence measurement simulation using the output from the calculation that produced Fig. 7. A coherence length of 5.8 mm is found at the SOA output and 9.7 mm at the filter output. The measured coherence length of the Santec HSL-2000 is 6.0 mm, a good match with the calculated output of the SOA.

The coherence calculation proceeds the way an experiment would. First simulated pulses are obtained. They are interfered over a range of time delays and the RF beat on a photodiode between DC and 60 MHz is calculated. This is wide enough that the simulated measurement produces the actual laser-limited coherence length, not being affected by a virtual instrumentation limit.

This and [12] are, to our knowledge, are the first coherence length calculations for a rapidly swept laser source appearing in the literature. We qualify that statement by pointing out that the "instantaneous linewidth" has been calculated in the case of an FDML laser [27], although no conversion to coherence length was provided. This kind of calculation will be of tremendous value in designing new lasers, given the central importance of coherence length on OCT system performance.

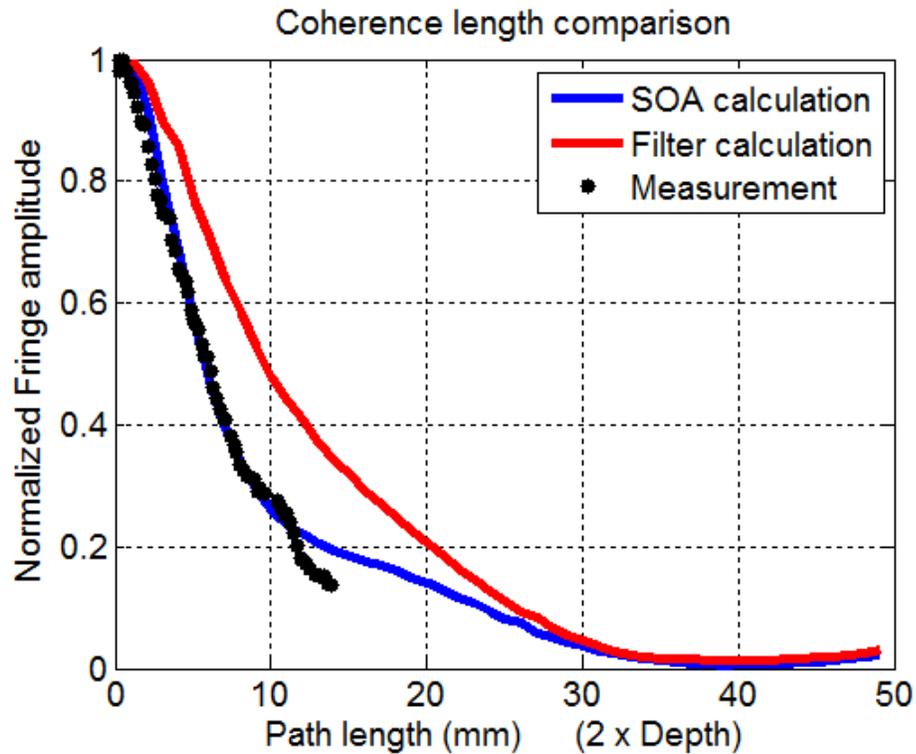
Fig. 8. Coherence experiment simulation using laser model.

Similar laser designs where the scanning mechanism was built to increase the cavity length in proportion to wavelength have been shown to have longer coherence length [20,21]. These are "mode-hop free" designs where the cavity length is always $m\lambda$, where $m$ is a fixed integer [25]. This condition, for a swept source, is the same as saying the Doppler shift from the cavity length change is the same as the required frequency hop determined by the tuning rate. In that case, no frequency shift from the gain medium is required and the laser can operate without pulsation. No pulses mean little chirp and longer coherence. We tested this idea by running the model over a range of Doppler shifts. Fig. 9 shows that this idea is correct, with a sharp peak in coherence when the Doppler shift matches the required frequency hop for the chosen tuning rate.

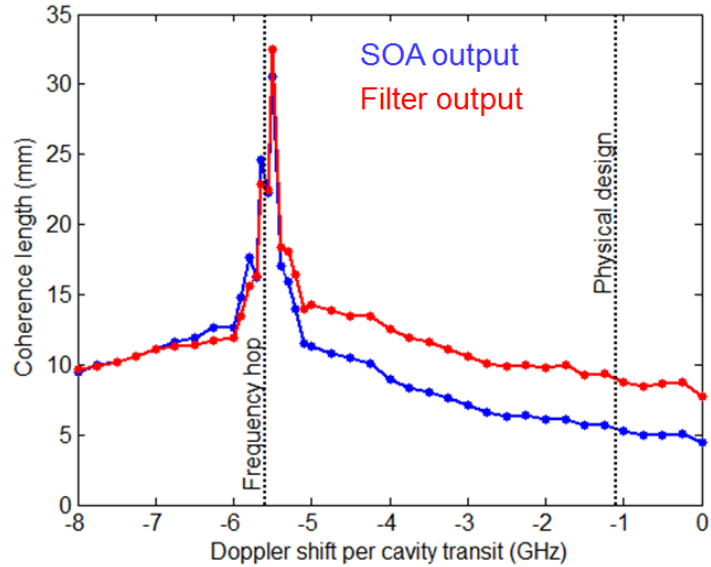

Fig. 9. Coherence calculations assuming other Doppler shift values induced by swept cavity length change.

**7. Summary**

Theory and experiment both show that many types of swept lasers, including the spinning polygon type studied here, are pulsed.   The emitted pulses modulate the gain medium in such a way as to promote short to long tuning.  Instead of new wavelengths being built up from spontaneous emission, each pulse hops to a longer wavelength by nonlinear means, tracking the swept optical filter.  This mechanism produces a low RIN laser with speed and coherence lengths suitable for a wide range of optical coherence tomography applications.

   The theoretical model can make many detailed predictions about pulsation and coherence.  In addition to the example shown here, similar calculations can predict the behavior of short cavity swept sources [12-14] and FDML lasers [28,29].  An experimental technique, the spectrogram, was developed as a method to track evolution of the laser operation during the wavelength sweep.  The method is potentially useful in the characterization of other swept sources.

   The theoretical understanding of swept lasers is important in allowing further commercial development of this technology and in enabling exploration of new designs, leading to performance improvements and more cost effective production.